\pdfoutput=1

\documentclass[apj]{emulateapj}
\usepackage{apjfonts}
\usepackage[bookmarks=true,colorlinks=true,citecolor=blue,linkcolor=magenta,urlcolor=cyan]{hyperref}

\usepackage{natbib}
\usepackage{amsmath}
\usepackage{amssymb}
\usepackage{subfigure}
\usepackage{url}
\usepackage{CJK}

\renewcommand{\vec}[1]{ {\mathbf #1} }

\newcommand{\Fig}{{Figure}}

\shorttitle{Data-Driven MHD modeling of AR 12192}
\shortauthors{Jiang et al.}

\begin{document}
\begin{CJK*}{UTF8}{gbsn}

  \title{How did a Major Confined Flare Occur in Super Solar Active Region
    12192?}

\author{
  Chaowei Jiang \altaffilmark{1,2},
  S.~T. Wu \altaffilmark{2},
   Vasyl Yurchyshyn \altaffilmark{3,5},
   Haiming Wang \altaffilmark{3,4},
  Xueshang Feng \altaffilmark{1},
  Qiang Hu \altaffilmark{2}}

\altaffiltext{1}{SIGMA Weather Group, State Key Laboratory for Space
  Weather, National Space Science Center, Chinese
  Academy of Sciences, Beijing 100190}

\altaffiltext{2}{Center for Space Plasma and Aeronomic Research, The
  University of Alabama in Huntsville, Huntsville, AL 35899, USA}

\altaffiltext{3}{Big Bear Solar Observatory, New Jersey Institute of
  Technology, 40386 North Shore Lane, Big Bear City, CA 92314, USA}

\altaffiltext{4}{Center for Solar and Terrestrial Research, New Jersey
  Institute of Technology, Newark, NJ 07102, USA}

\altaffiltext{5}{Korea Astronomy and Space Science Institute, 776
  Daedeok-daero, Yuseong-gu, Daejeon, 305-348, Korea}

\email{cwjiang@spaceweather.ac.cn}

\begin{abstract}
  We study the physical mechanism of a major X-class solar flare that
  occurred in the super NOAA active region (AR)~12192 using a
  data-driven numerical magnetohydrodynamic (MHD) modeling
  complemented with observations. With the evolving magnetic fields
  observed at the solar surface as bottom boundary input, we drive an
  MHD system to evolve self-consistently in correspondence with the
  realistic coronal evolution. During a two-day time interval, the
  modeled coronal field has been slowly stressed by the photospheric
  field evolution, which gradually created a large-scale coronal
  current sheet, i.e., a narrow layer with intense current, in the
  core of the AR. The current layer was successively enhanced until it
  became so thin that a tether-cutting reconnection between the
  sheared magnetic arcades was set in, which led to a flare. The
  modeled reconnecting field lines and their footpoints match well the
  observed hot flaring loops and the flare ribbons, respectively,
  suggesting that the model has successfully ``reproduced'' the
  macroscopic magnetic process of the flare. In particular, with
  simulation, we explained why this event is a confined eruption--the
  consequent of the reconnection is the shared arcade instead of a
  newly formed flux rope.  We also found much weaker magnetic
  implosion effect comparing to many other X-class flares.
\end{abstract}


\section{Introduction}
\label{sec:intro}

Solar flares are sudden release of excess magnetic energy in the solar
corona, a plasma environment dominated by the magnetic
field~\citep{Shibata2011}. Magnetic reconnection is believed to be the
central mechanism that converts free magnetic energy into radiation,
energetic particle acceleration, and kinetic energy of
plasma~\citep{Forbes2006}. Consequently, revealing the magnetic
structures associated with reconnection and their evolution during
flares is essential for understanding of the flare
dynamics~\citep{Priest2002}.

Due to the lack of direct measurements of coronal magnetic fields, it
is a prevailing way to postulate the flare magnetic evolution from the
observed variations of flare plasma emission. This is because the
plasma emission can reflects the geometry of the invisible magnetic
field, as in most part of the corona, the plasma is ``frozen'' with
the magnetic fields. Early studies of typical eruptive flares have
converged to a standard flare model~\citep[CSHKP,][]{Carmichael1964,
  Sturrock1966, Hirayama1974, Kopp1976}, which describes the essence
of flare physics.  The standard model mainly concerns a magnetically
bipolar source region, the simplest form of solar active regions
(ARs), proposing that a twisted magnetic flux rope (corresponding to a
filament) rises above the polarity inversion line (PIL), stretches the
overlying closed field lines (manifested as coronal loop expansion),
and produces a vertical current sheet (CS) underneath where
reconnection sets in and results in two parallel chromospheric flare
ribbons on both sides of the PIL. The flare ribbons are suggested as
the footprints of the reconnecting field lines. They gradually move
apart from one another as the reconnection goes on. Meanwhile, the
ejecting flux rope eventually travels into solar wind as being a
coronal mass ejection (CME), leaving behind bright flaring loops that
correspond to the re-closed magnetic arcades after the
reconnection. Such dynamic picture inferred from observations is
usually represented by simple cartoons\footnote{see an archive of such
cartoons on \url{http://solarmuri.ssl.berkeley.edu/~hhudson/cartoons/}}.

Recent observations with high spatial/time resolution and
multi-wavelength imagers show that numerous solar flares are
characterized by complex processes that are not present in the
standard model, such as multi-stage and multi-place of filament
ejections in the same event~\citep[e.g.,][]{LiuC2009, Schrijver2011a,
  ShenY2012, Romano2015}, escape of homologous flux
ropes~\citep[e.g.,][]{LiT2013a}, slipping motions of flare
loops~\citep[e.g.,][]{Aulanier2007, LiT2015, Dudik2016, GouT2016},
flare ribbons of unusual shapes~\citep[e.g., quasi-circular and even
tri-linear shapes,][]{Masson2009, WangH2012, WangH2014}, multiple
ribbons like remote flare ribbons distinct from the eruptive core
site~\citep[or the secondary ribbon, e.g.,][]{ZhangJ2014}, the EUV
late phase after the main (impulsive) phase in certain
flares~\citep{Woods2011, Dai2013, LiuK2013} and etc.
There are also flares without CMEs, which are usually called
confined flares.  Some confined flares occur with filament
eruptions but failing to escape their overlying field~\citep{Ji2003,
  Torok2005, GuoY2010}. The others, simply without any eruption, are the most hard
to interpret solely from observations, because very small changes of
the coronal configuration can be detected in these
flares~\citep[e.g.,][]{Jiang2012c, Dalmasse2015}.


To understand the mechanisms of the various complex or atypical flares
requires us to characterize the realistic magnetic configurations and
their evolution associated with flares. Also, from the point of view
of prediction of space weather, which is heavily influenced by solar
eruptions, a much more accurate understanding and reproducing of the
eruption process beyond the standard or theory model is strongly
required.  Existing techniques to this end include static non-linear
force-free field (NLFFF) reconstruction~\citep[see review papers
by][]{Wiegelmann2012solar, Regnier2013}, data-constrained/driven
magneto-frictional (MF) evolution
method~\citep[e.g.,][]{Cheung2012,Yeates2014,Savcheva2015,
  Fisher2015}, data-constrained magnetohydrodynamic (MHD)
simulations~\citep[e.g.,][]{Jiang2012c, Jiang2013MHD, Kliem2013,
  Amari2014nat, Inoue2014, Inoue2015}, and more generally, the
data-driven MHD simulations~\citep[e.g.,][]{Wu2006}.

Among the available techniques, the NLFFF reconstruction is used most
frequently because a variety of approaches and codes for solving NLFFF
have been developed within in a relatively long
history~\citep[e.g.,][]{Grad1958, Sakurai1981,Yang1986,Wu1990,
  Amari1997, Yan2000, He2006, Wiegelmann2006,
  Wheatland2006,Valori2010,Jiang2012apj}, and prove to be successful
for studying snapshots of the coronal fields before and after
flares. Signature of flare mechanism can usually be suggested from
analysis of the pre-flare magnetic fields. For example, through
studying the magnetic topology, critical magnetic structures relevant
to flares, such as magnetic flux rope, magnetic null points, bald
patch~\citep{Titov1993}, and quasi-separatrix
layers~\citep[QSLs,][]{Demoulin1996, Titov2002} could be revealed.
However, analyzing the pre-flare fields cannot tell directly why and
how the flares occur.  The lack of dynamics is a major limitation of
NLFFF reconstruction, which cannot be used for ``reconstruction'' of
magnetic field during flares. The limitation exists similarly in the
MF methods. Although in such methods, a dynamic velocity is included
for making the magnetic field ``evolves'', this velocity is actually
pseudo since it is determined only by the Lorentz force (i.e., the
veloctiy $\vec v = \vec J\times \vec B/\nu$ where $\nu$ is the
frictional coefficient) while the inertia and pressure of the plasma
is neglected~\citep{Yang1986}. As a result, the MF approach are still
limited for the quasi-static evolution phase of the corona field.
When used for modeling the magnetic field evolution, the MF method is
essentially similar to the way using a time-sequence of NLFFF or MHD
models reconstructed independently from a series of vector magnetogram
along time to mimic the coronal evolution, although in the MF method,
the magnetic fields for each time snapshot are treated to be dependent
on its preceding one.  It is still problematic to use the MF method to
simulate the flare and eruption phase in which the plasma is in
extremely dynamic evolution and often associated with magnetic
reconnections, although such way has been used in analysis of
evolution of flare ribbons~\citep[e.g.,][]{Savcheva2015, Savcheva2016,
  Janvier2016}.  The data-driven MHD model of \citet{Wu2006}, however,
just uses the line-of-sight magnetograms, and thus the
non-potentiality of the coronal field cannot be fully recovered. There
are models~\citep{Kliem2013, Jiang2013MHD, Amari2014nat, Inoue2014}
using the NLFFF reconstructed or MF calculated coronal field
immediately preceding eruption (thus the unstable nature of the field
has already well developed) as the initial condition for MHD
simulation, which prove to be able to reproduce the fast dynamic phase
of the erupting field~\citep[e.g.,][]{Jiang2013MHD}. However, these
kinds of simulations do not self-consistently show how the
pre-eruptive field is formed and disrupted, and thus may not be used
to identify the true triggering mechanism. Also, such kind of models
might not be able to reproduce confined flare, which is not likely
triggered by the large-scale instability of the pre-flare magnetic
field.

To self-consistently and realistically simulate the coronal evolution
from its pre-flare to flare phases, we have developed a new
data-driven 3D MHD AR evolution (DARE) model. The DARE model is based
on the full MHD equation with its lower boundary driven directly by
the solar vector magnetograms, which is unique among all the
aforementioned models that attempt to simulate the realistic coronal
magnetic evolution. In the first application of this
model~\citep{Jiang2016NC}, we modeled the evolution of a complex
multi-polar AR with flux emergence over two days leading to an
eruptive, also atypical flare. The simulation reasonably recreated the
whole process from a long quasi-static evolution to the eruptive stage
of extreme dynamics. It was shown that the field morphology resembles
the sequence of the corresponding EUV images from SDO/AIA for such
process, in addition to the successful match of the timing of the
flare onset.

In this paper, we propose to use the DARE model to study a distinctly
different event, a confined major X-class flare occurred in the super
NOAA AR~12192 on October 2014. This AR is ``super'' because of its
size, which is the largest of all ARs during the last 25 years. It
produced a series of X-class flares without eruptions, and the
strongest one in the series reaches X3.1, which sets a record in the
flare energy for CMEless events~\citep{Thalmann2015} since the
confined flares ever observed were predominantly below
X-class~\citep{Yashiro2006}. A series of studies have been inspired to
explain why these extremely powerful flares did not lead to
eruptions~\citep{JingJ2015, SunX2015, ChenH2015, Inoue2016}. Here we
are curious why and how did these flares occur, or more specifically,
what is the evolution of the magnetic fields underlying these
non-eruptive flares in AR~12192? We attempt to answer this question by
simulating the coronal magnetic field evolution of the AR leading to
the X3.1 flare.  The rest of this paper is organized as follows. The
flare event to be studied are described in Section~\ref{sec:event}.
Then the DARE model is briefly presented in Section~\ref{sec:model}.
Results are given in Section~\ref{sec:res} and finally discussions in
Section~\ref{sec:con}.



\begin{figure}[htbp]
  \centering
  \includegraphics[width=0.45\textwidth]{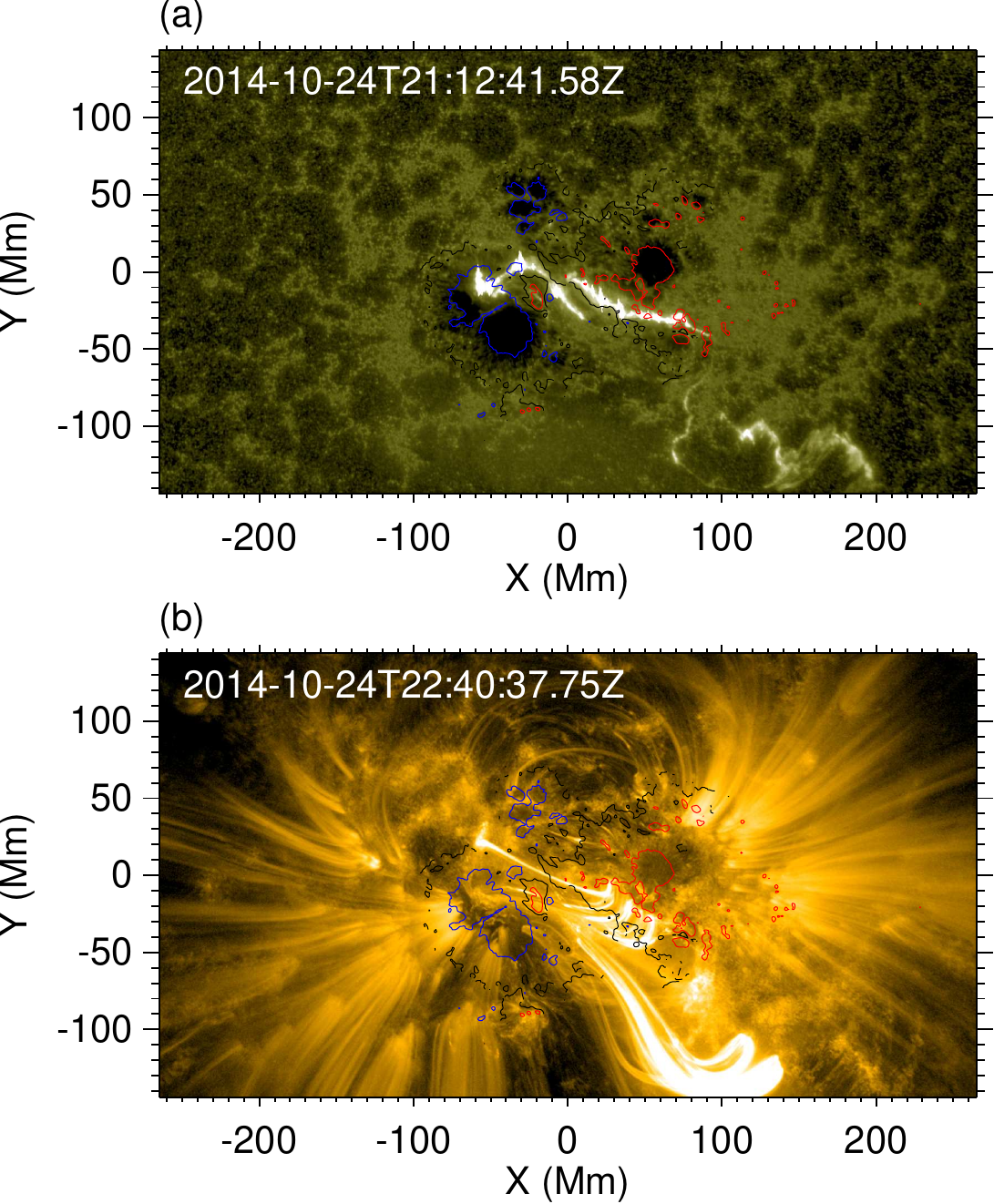}
  \caption{Overview of the X3.1 flare occurred in AR~12192. (a) AIA
    1600~{\AA} image of the flare ribbon and (b) AIA 171~{\AA} image
    of the post-flare loop. Both images are CEA re-mapped with field
    of view identical to that of the SHARP CEA magnetogram for this
    AR. Overlaid contours indicate the vertical component ($B_{z}$) of
    the photospheric magnetic field at the same time (red for
    $1000$~G, blue for $-1000$~G, and black for the PIL with the total
    field strength above $200$~G).}
  \label{fig:fulldisk}
\end{figure}

\section{Event}
\label{sec:event}

Since overview of the AR~12192 and its unusual feature, that is,
extremely large size, rich of X-class flare but CME-poor, has been
described well in the literature~\citep{Thalmann2015, SunX2015,
  ChenH2015, JingJ2015, ChenH2015}, we focus on the X3.1 event and the
relevant information with our modeling. In the period of our interest
from 2014 October 23 to 24, AR~12192 is close to the central
meridian. Two major sunspots are well separated by a distance of roughly
$100$~Mm (see \Fig~\ref{fig:fulldisk}a). The target flare occurred
around 21:00~UT, October 24, and it lasts for an unusual long
duration of more than one hour with the GOES X-ray flux above X
class. Preceding the major flare are relatively small ones of C- and
M-class with shorter durations. When inspecting the SDO/AIA images
(for instance, \Fig~\ref{fig:fulldisk}b) one only see a series of
brightening of coronal loops without much changes in their shape. The
flare loops are seen connecting the boundaries of the strong magnetic
polarities. Interestingly, there is also a set of rather long loops
with remote connection to the southwest corner of the field of view as
shown \citep[e.g.,][]{SunX2015, ChenH2015}.  Chromospheric ribbons of
the flare (\Fig~\ref{fig:fulldisk}a) consist of mainly two
bands on both sides of the central part of the PIL. Distinct from
typical two-ribbon flares, these two ribbons showed barely separation motion.
As such,  the coronal configuration changes are not easy to interpret from
these EUV observations.

\section{The DARE Model}
\label{sec:model}

In the DARE model~\citep{Jiang2016NC}, we used the solar surface
magnetic field data from the SDO/HMI~\citep{Schou2012HMI}, in
particular, the Space weather HMI Active Region Patches (SHARP) vector
magnetogram data series \citep{Hoeksema2014, Bobra2014}. With the
cadence of 12~min and the spatial resolution of 1~arcsec, the SHARP
data are adequate to track a relatively long-term evolution (hours to
days) of magnetic structures of the typical AR scale.  To setup the
model, we considered a local Cartesian coordinate system with its
origin at the surface center of the AR, which is defined in the
cylindrical equal area (CEA) re-mapped SHARP magnetogram. The 3D
computational volume extends approximately $[-450, 450]$~Mm in both
$x$ and $y$ axes and 900~Mm in $z$ axis, which is sufficient to
include the large-scale magnetic field related with the
flare. Furthermore, all the external boundaries (except the bottom
surface) are non-reflective. We note that here the AR spans $\sim
60^{\circ}$ of the solar sphere, much larger than typical ARs, and the
curvature of the associated area should not be ignored. Thus the
current modelling in Cartesian box might not appropriately
characterize the geometry for the AR, and we keep in mind the possible
influence in discussing our modeled results.

Based on the HMI vector magnetogram, we first constructed an
approximately force-free coronal magnetic field~\citep{Jiang2013NLFFF}
corresponding to a pre-flare instance of 00:00~UT on 2014 October~23.
Then, with this field as the initial condition, we numerically solved
the full set of time-dependent, 3D MHD equations in the modeling
volume. The bottom boundary of the model is assumed as being the
coronal base, thus the magnetic field measured on the photosphere is
used as a reasonable approximation of the field at the coronal
base. Then the evolving solar surface magnetic fields from observation
provide the time-dependent bottom boundary conditions for the
simulation domain. We smoothed the original SHARP data before
inputting them into the numerical model.  This is necessary since the
magnetic structures are broadened from the photosphere to the coronal
base. We simulated such broadening using Gauss smoothing of the data
with Gaussian window of $\sigma=2$~arcsec as suggested
by~\citet{Yamamoto2012}. We further smoothed the data in time with
Gaussian window of $\sigma=4\times12$~min to remove short-term
temporal oscillations and mitigate the problem of data spikes due to
bad pixels. Interpolation in time was employed to fill small data gaps
in the two days of observation.

In addition to the magnetic field, we also need to give a model of
plasma in the computation. Here, the plasma is initialized in a
hydrostatic, isothermal state with $T=10^{6}$~K (sound speed
$c_{S}=128$~km~s$^{-1}$) in solar gravity. Its density is configured
to make the plasma $\beta$ as small as $2\times 10^{-3}$ (the maximal
Alfv{\'e}n $v_{\rm A}$ is $4$~Mm~s$^{-1}$) to mimic the coronal
low-$\beta$ and highly tenuous conditions. The plasma thermodynamics
are simplified as an adiabatic energy equation since we focus on the
evolution of the coronal magnetic field. No explicit resistivity is
included in the magnetic induction equation, and magnetic reconnection
is still allowed due to numerical diffusion if any CS forms and
becomes thin enough with thickness close to the grid resolution (i.e.,
the smallest grid). A small kinematic viscosity $\nu$ is used with its
value corresponding to the viscous diffusion time as $\sim 10^{2}$ of
the Alfv{\'e}n time in strong-field regions. This is usually necessary
for the sake of numerical stability in the long term computation. The
units of length and time in the model are $L=23$~Mm (approximately
32~arcsec on the Sun) and $\tau=L/c_{S}=180$~s, respectively.

Solution of the MHD equations is implemented by an advanced space-time
high-accuracy scheme~\citep[AMR--CESE--MHD,][]{Jiang2010}. We use a
non-uniform grid based on the magnetic flux distribution for the sake
of saving computational resources. The smallest grid $\Delta x =
\Delta y = \Delta z=1.4$~Mm (approximately 2~arcsec on the Sun) is
made around the AR core region (approximately $[-100,100]\times
[-100,100]\times [0,100]$~Mm$^{3}$), where the magnetic fields are
strong and evolve actively. Grid size is increased gradually to
$4\Delta x$ near the side and top boundaries.

To further save the computing time, the cadence of the input HMI data
into the MHD model was increased by $20$ times. By this, the model run
of a realistic two-day AR evolution can be finished within about ten
hours of wall time when parallelized with a medium number (for
example, a hundred) of CPUs (3~GHz). Compressing of the time in HMI
data is justified by the fact that the speed of photospheric flows as
measured from the photospheric field evolution is about $0.1\sim 1$
km~s$^{-1}$~\citep{Welsch2004, Liu2012}. So in our model settings, the
evolution speed of the boundary field, even enhanced by a factor of
20, is still sufficiently small compared with the coronal Alfv{\'e}n
speed ($\sim$Mm~s$^{-1}$ ), and the basic reaction of the coronal
field to the bottom changes should not be affected.  As a result, one
hour in the HMI data equals one $\tau$ in the simulation.  When
comparing the simulation with the observations of the corona, such
scaling of time (a $\tau$ equals an hour) also applies to the
quasi-static evolution phase without major flares or eruptions. This
is because in such phase, the coronal evolves in the same pace as the
boundary, since any change in the bottom boundary is reflected almost
instantly in the corona, which reaches its equilibrium very fast. But
this is not justified for the dynamic phases with major eruptions or
flares, in which the evolution of the corona is determined by itself
rather than by the boundary. Thus, the time unit should be the
original one (a $\tau$ equals $180$~s) in the flare phase.

Coupling of the coronal evolution with the continuous change of the
surface magnetic field (i.e., the HMI data) is implemented by a
time-dependent bottom boundary condition using the
projected-characteristic method. Based on the wave-decomposition
principle of the full MHD system~\citep{Nakagawa1987, Wu2006}, the
method can naturally mimic the transferring of magnetic energy and
helicity to the corona from below~\citep{Wu2006} by self-consistently
calculating the surface flow field~\citep{Wang2008ApJ}, which
otherwise would have to be derived by local correlation tracking or
similar techniques~\citep{Welsch2004, Schuck2008}. As in our settings,
the cadence of HMI data is $\tau/5=36$~s. However, the time step in
the numerical model is set as $\Delta t = {\rm CFL} \min(\Delta
x/v_{\rm A}) = 0.2$~s according to the Courant--Friedrichs--Lewy (CFL)
stability condition~\citep{courant1967partial} with a CFL number of
0.5. We thus linearly interpolate the HMI data in time to produce a
data set with cadence matching the time step of the MHD model.

We followed the evolution of the MHD system for two days from 00:00~UT
of October 23 ($t=0$) to 00:00~UT of October 25 ($t=48$).

\begin{figure*}[htbp]
  \centering
  \includegraphics[width=0.9\textwidth]{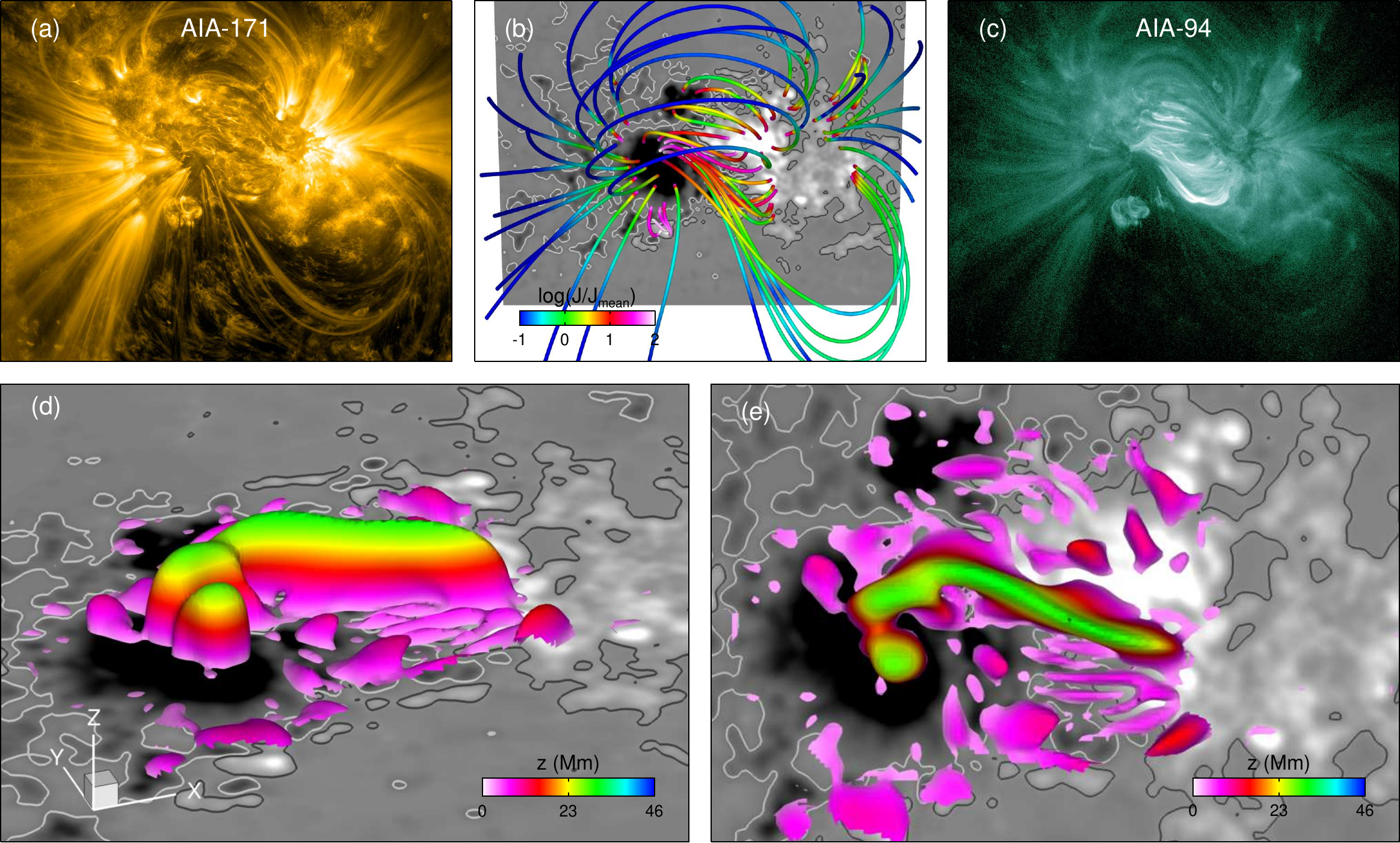}
  \caption{AIA images and the simulated magnetic configuration of
    AR~12192 at 00:00 UT of 2014 October 23 (i.e., simulation time
    $t=0$). (a) AIA 171~{\AA} image. (b) Selected magnetic field lines
    of the MHD simulations. The background shows map of the
    photospheric $B_z$ (saturated at $\pm 1000$~G). Contour lines
    (black and white) are also shown for $\pm 100$~G.  The color of
    the plotted field lines denotes the strength of the associated
    electric current density $J$ (scaled by the mean value $J_{\rm
      mean}$ of the entire model volume). (c) AIA 94~{\AA} image. (d)
    The side view of the volume with the current density higher than
    20 times the mean value. The color denotes the height $z$ from the
    bottom boundary. (e) The top view of the same current volume shown
    in (c). Contour lines (black and white) are also shown for $\pm
    100$~G.}
  \label{fig:initial}
\end{figure*}

\section{Results}
\label{sec:res}

\subsection{The initial state}
\label{sec:3.1}

In \Fig~\ref{fig:initial} we show the magnetic configuration derived
from the near force-free model for the initial state ($t=0$) of the
MHD simulation. On the large scales, the AR exhibits a bi-polar
magnetic configuration consisting of two main sunspots with a
relatively strong-sheared core fields embedded in a less-sheared
envelope fields. The core fields carry relatively much stronger
electric current (e.g., current density higher than 20 times of the
average value of the whole model box, see \Fig~\ref{fig:initial}b, d
and e) that is concentrated within a narrow vertical layer roughly
along the central part of the PIL separating the two major
polarities. Such an association of intense current layer with PIL of
AR core might be common for flare-productive
ARs~\citep[e.g.,][]{SunX2015}. Here the current layer extends from the
bottom to a relatively large height ($\sim 35$~Mm, see
\Fig~\ref{fig:initial}d). From a visual comparison with the EUV images
(\Fig~\ref{fig:initial}a, b and c), the simulated magnetic field lines
show good agreement with the observed coronal loops. In particular, it
can be seen that the weakly sheared envelope fields resemble the long
cool loops imaged in the AIA 171~{\AA} channel (about 1~MK), and the
strongly sheared core fields resemble the short hot loops imaged in
the AIA 94~{\AA} channel (about 6~MK). This is likely due to heating
by dissipation of the strong current in the core region, making the
plasma there hotter than the surroundings. In the following we show
how this coronal field evolved when driven by the photospheric
magnetic evolution.

\begin{figure*}[htbp]
  \centering
  \includegraphics[width=0.8\textwidth]{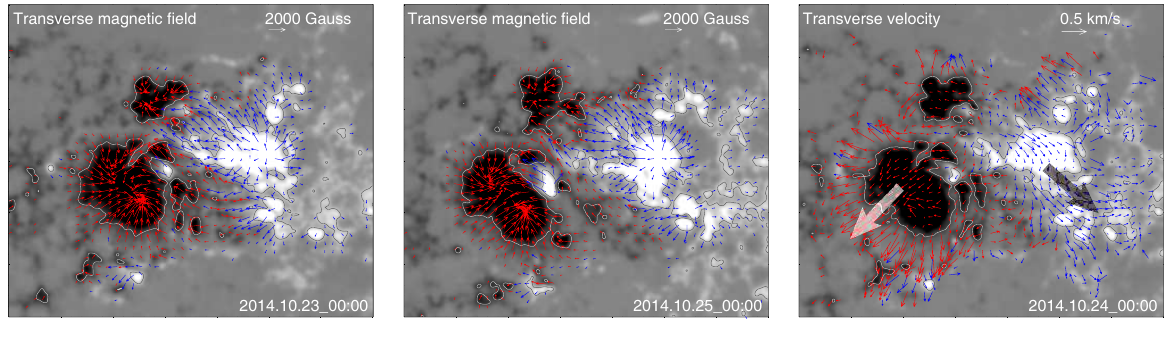}
  \caption{Left and middle panels are SDO/HMI vector magnetograms
    (left and middle) for AR~12192 recorded at two different
    times. The vertical component $B_z$ as shown saturated at $\pm
    1000$~G. Right panel are photospheric horizontal velocity derived
    using the DAVE4VM code, overlaid on the $B_z$ map.  The two big
    arrows illustrate the large-scale movement of the magnetic
    polarities. {\it See also animation of the evolution of the vector
      magnetograms during two days of interest.}}
  \label{fig:HMIVM}
\end{figure*}

\begin{figure*}[htbp]
  \centering
  \includegraphics[width=0.8\textwidth]{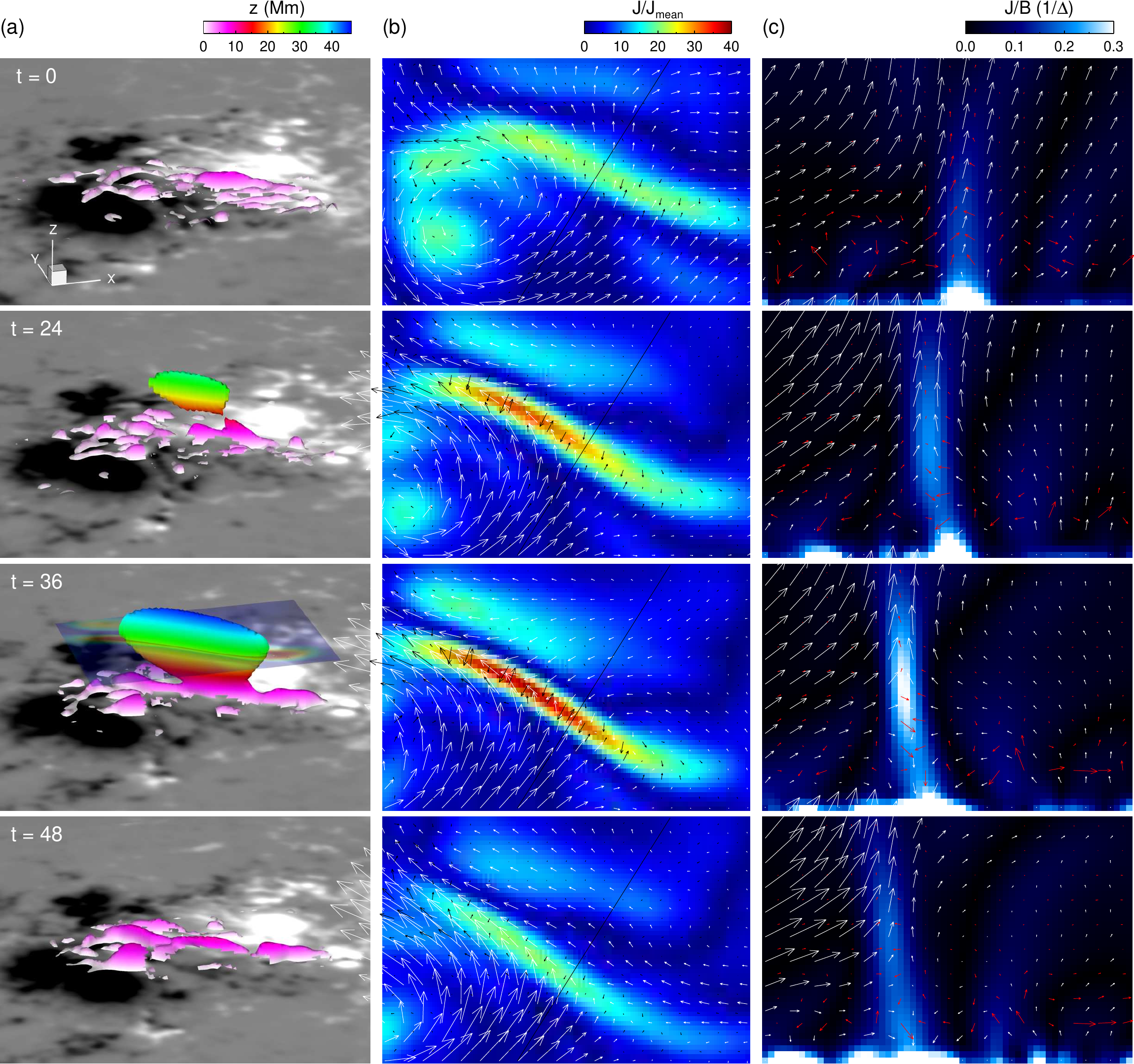}
  \caption{Evolution of the electric current in the MHD model. (a) 3D
    shape of the CS (defined in the text) at four different times from
    the beginning to the end of the simulation. The color denotes the
    height from the bottom surface, which is shown by the photospheric
    $B_z$ map. (b) A horizontal slice of the volume at $z=28$~Mm. Its
    position is indicated by the transparent image shown in $t=36$ of
    (a).  The color shows electric current density $J$ (scaled by the
    average value $J_{\rm mean}$ at $t=0$), and the arrows show plasma
    velocity vectors (white arrows) as well as Lorentz force vector
    (black arrows). The time is the same as (a) from top to
    bottom. (c) A vertical slice whose horizontal location is denoted
    by the black inclined line in (b). Vertically, it extends from the
    bottom to $z=57$~Mm. The color shows the value of $J/B$ and the
    arrows represent the velocity (white) and Lorentz force
    (red). Note that in (b) and (c) each pixel represents a
    computational grid. {\it An animation for the 3D CS evolution and
      a horizontal slice is provided, in which the time step is
      $\tau/5$.}}
  \label{fig:slice}
\end{figure*}

\begin{figure}[htbp]
  \centering
  \includegraphics[width=0.4\textwidth]{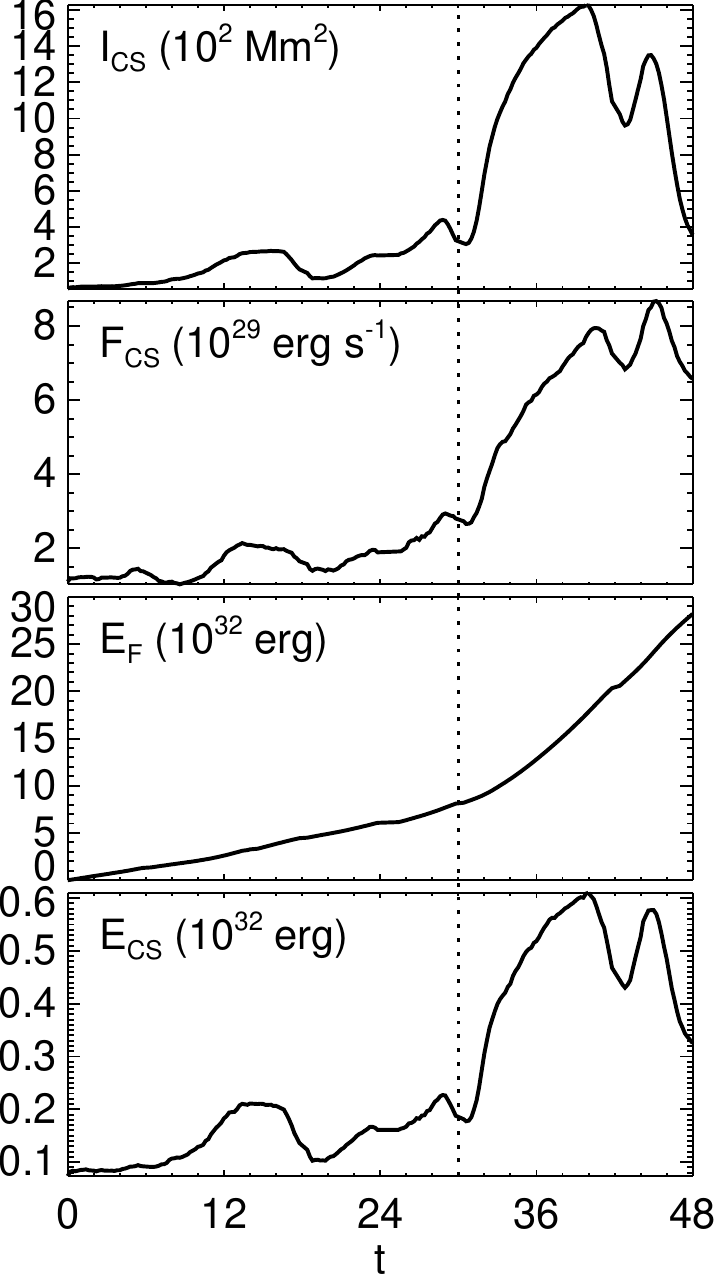}
  \caption{Temporal evolution of different parameters (defined in the
    text) associated with the CS. (a) The intensity of the CS. (b) Rate
    of magnetic energy injection into the CS. (c) Total magnetic
    energy injected into the CS with time. (d) Total magnetic energy
    reserved in the CS volume. The dashed line denotes $t=30$.}
  \label{fig:CS_lineplot}
\end{figure}

\subsection{Dynamic evolution}
\label{sec:3.2}

\Fig~\ref{fig:HMIVM} and its supplementary animation show how the
photospheric magnetic field evolved over the two day time
period. Continuous movements of the polarities can be seen, which
could stress the coronal field. For instance, a horizontal flow map
derived for the time moment of 00:00~UT on 2014 October 24 using the
DAVE4VM method \citep{Schuck2008} demonstrates clearly a diverging
motion of the two main sunspots. Driven by the magnetic field
evolution at the bottom, the MHD system continuously evolved in
response. The basic configuration of the coronal field shows no
significant changes when we trace the magnetic field lines (thus not
shown in the figures here), and the total kinetic energy maintains in
a rather low level (less than 0.5 percent of the total magnetic
energy) without significant variation in the whole time interval,
indicating that no eruption occurs in the simulation. However,
evolution of the distribution of electric current, particularly the
CS, a thin layer of intense current as defined below, provides
instructive information that might be associated with the flare
processes.

Here we use a way following~\citet{GibsonFan2006} to locate the CS.
It is defined as the volume in which the ratio of the current density
to the magnetic field strength, i.e., $J/B$, is greater than $C/\Delta x$, where
$\Delta x$ is the local grid size and the constant number $C \sim
1$. Such a definition is reasonable because $J/B \sim (\Delta
B/B)/\Delta x $, where $\Delta B$ denotes the gradient of magnetic
field vectors in adjacent grid points, and $\Delta B/B << 1$ in the
smooth region of magnetic field (except regions with $B\sim 0$, for
example, near magnetic null points). While a large value of $\Delta
B/B \sim 1$ indicates significant change of magnetic field (as large
as the local field strength) between adjacent grid points. This means
that the magnetic field vectors of distinctly different directions are
squeezed extremely close to each other, forming a narrow interface
with strong current, which is a CS in the context of our numerical
model. When the gradient of magnetic field vectors across the CS is
steepened sufficiently, numerical diffusion will take effect and
``reconnection'' could occur in the MHD model. In other words, for
such case, the inversely-directed magnetic components on both sides of
the CS are brought so close to each other that they ``merge'' within
the CS, resulting in new connections of the corresponding magnetic
field lines, and thus related to topological changes of the field. By
inspecting the value of $J/B$, we find a suitable value of $C=0.2$
which can well exclude the region of weak current, and such definition
gives the width of the CS of $\sim 3\Delta x$. It should be noted here
the CS is different from that in theoretical view, i.e., an infinitely
thin or simply a 2D surface rather than a finite volume.

The 3D shapes of the CS at different times are shown in
\Fig~\ref{fig:slice}a (and also in its supplementary animation).  A
horizontal slice of the volume is shown in \Fig~\ref{fig:slice}b, and
a vertical slice in \Fig~\ref{fig:slice}c.  Initially ($t=0$) the CS
did not yet form, thus there are only small-scale structures near the
bottom surface.  Then, successive formation, expansion and shrinkage
of the CS volume are seen. From the plasma flows and the Lorentz force
vectors around the CS volume, we can see that the initial current
layer is gradually squeezed from both two sides by the magnetic
stress. Consequently it becomes thinner and more intense, leading to
increasing of the value $J/B$ and thus the formation of the CS as we
defined ($J/B > 0.2\Delta x$). At the time of $t=36$ (close to the
peak stage of the CS development, see below), the CS extends from the
bottom of the simulated domain up to heights of 50~Mm. With the slow
evolution of the basic configuration, the CS also moves slowly from
north to south, but overall its location is roughly the same in whole
period.  Interestingly, the pattern of reconnection-like plasma flow,
i.e., horizontal inflow at both sides of the CS and vertical outflow
to up and down, can be seen (\Fig~\ref{fig:slice}b and c, see $t=24$
and $36$, for example), suggesting that reconnection might occur in
the simulation.

We then calculated the following parameters to characterize the CS
evolution and to see whether reconnection occurred within the CS:
\begin{enumerate}
\item Since in the evolution process the CS grows and decays
  dynamically, for each time snapshot, we integrate $J/B$ for the
  full volume $V$ of the CS as defined above, which gives a value
  $I_{\rm CS}=\int J/B dV$ in unit of area, and can be used to quantify the size or
  intensity of the evolving CS;
\item Furthermore, we calculated the rate of the magnetic energy
  injection into the CS by $F_{\rm CS} = -\int_{S} \vec P dS = -\int_{V}
  \nabla \cdot \vec P dV$, where $\vec P = \frac{1}{4\pi}\vec
  B\times(\vec v\times \vec B)$ is the Poynting flux vector with $\vec v$
  as the plasma velocity, and $S$ is the surface area of the CS volume $V$;
\item The total magnetic energy injected into the CS, i.e., $E_F =
  \int_{0}^t F_{\rm CS} dt$ at time $t$;
\item As here the CS is not a 2D surface but has a finite volume, we
  would like to know the magnetic energy ($E_{\rm CS}=\int_{V}
  \frac{B^2}{8\pi}dV$) in the CS volume. So the happening of reconnection
  can be indicated if the magnetic energy stored in this volume is less than
  those injected into it.
\end{enumerate}

The results are shown in \Fig~\ref{fig:CS_lineplot}. If omitting the
small fluctuations of the CS size profile during the whole process, we
find that its evolution consists of two phases of distinct behaviors:
in the first phase from $t=0$ to approximately $30$, it keeps a
relatively small value ($\sim 200$~Mm$^2$); in the second phase from $t=30$ to the end of
the simulation, it first increases impulsively and reaches the peak
value of 1600~Mm$^2$ at nearly $t=40$, and then decreases rapidly to a value similar to that in the
first phase. The evolution of the rate of magnetic energy injected
into the CS, $F_{\rm CS}$, shows a similar trend.
In the later phase (i.e., from $t=30$ to the end), the total magnetic
energy input is about $10^{33}$~erg, while the magnetic energy in the
CS volume is smaller by nearly two orders in magnitude and can be
negligible. This indicates that the amount of the magnetic energy
injected into the CS is mostly released via reconnection (converted
into other forms), and the energy release rate can be approximated by
$F_{\rm CS}$. If we regard the impulsive increase and decrease of the
energy release rate in the CS as a simulated ``flare'' process, we
estimate the released energy by the flare is about $10^{33}$~erg and
the energy conversion rate is on the order of $10^{29}$~erg~s$^{-1}$,
and can reach an order higher in the peak time. For a reference, the
potential energy of this AR is approximately $1.5\times 10^{34}$~erg
during our studied period, over an order higher than typical-size ARs,
e.g., AR~11158~\citep{SunX2015}. On the other hand, the first phase
($t=0$ to $30$) can be regarded as a quasi-static evolution duration
for which the time unit $\tau$ can be scaled as being one hour, as
mentioned in Section~\ref{sec:model}.  This means that our simulated
flare ($t=30$) began $15$ hours ahead of the real X3.1 flare ($t\sim
45$). 
However, for the simulated flare phase,
the time duration of nearly 20~$\tau$, and thus equaling one hour (since $\tau=180$~s),
is close to the real one that has
relatively long X-ray duration of about one hour.

The above analysis of the CS evolution based on the modeled results
suggests a reasonable picture of how the real flare was produced:
photospheric field evolution stressed the coronal field and built up a
large-scale CS in the core region, then magnetic reconnection was
triggered immediately and resulted in impulsive release of magnetic
energy. As the kinetic energy is very low even during the impulsive
phase of energy release, the most of the released magnetic energy
should be converted into energy of accelerated electrons (ions also
possible), and subsequently in the form of radiation at various
wavelengths.
However, our simulation cannot reproduce this process because we did
not include the related physics in the model. Nevertheless, the
modeled results are in an agreement with the fact that this was a
non-eruptive flare, during which no significant disruption of the
coronal field was observed.

It is worthy noting that there were small fluctuations, i.e., short
episodes of relatively small-scale CS formation and dissipation before
the major one.  These fluctuations reflect the energy build up and
might be the simulation counterparts of the small flares that occurred
before the main one (but a one-to-one correspondence was not
reproduced). In fact, it signifies that the magnetic field is stressed
and from time to time and the process is interrupted by episodic
small-scale reconnection. However, preceding to the major flare, the
CS is not large enough for the global-scale reconnection to set in.

\begin{figure*}[htbp]
  \centering
  \includegraphics[width=0.8\textwidth]{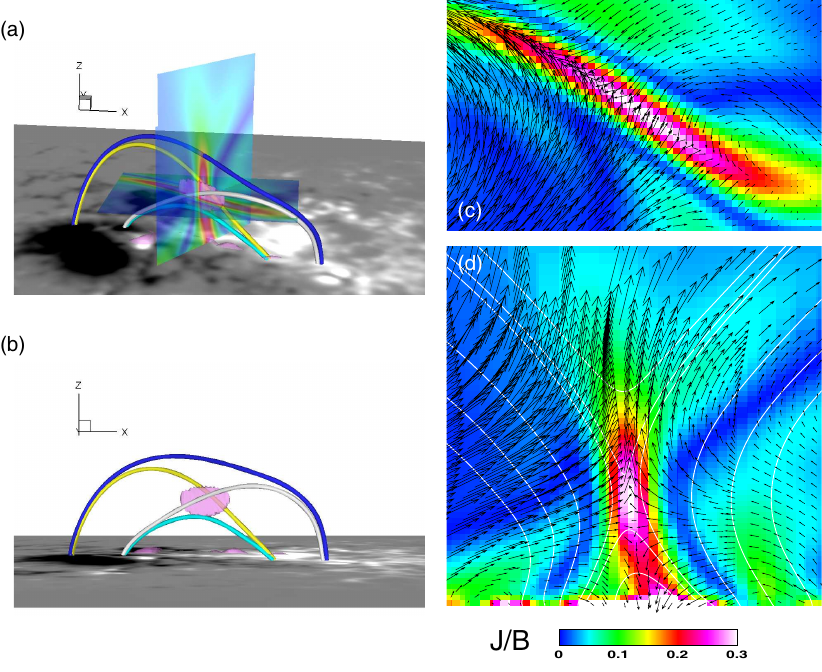}
  \caption{Illustration of the tether-cutting reconnection process in
    the model. (a) Sampled field lines and horizontal and vertical
    cross sections of the reconnection site.  The white and yellow
    curves represent the ``before'' flare field lines and blue and
    cyan curves show the new reconnected field lines. The pink colored
    area shows the iso-surface of $J/B=0.3/\Delta x$, which is
    sandwiched between the pre-reconnection field lines at their
    crossing point. (b) Side view of the sampled field lines.
    (c) and (d) The horizontal and vertical
    cross sections zoomed-in to show details of the $J/B$ (in unit of
    $1/\Delta x$) structures and plasma flow vectors. The vertical
    axis of in (c) is $z$ in range of $[0,92]$~Mm. The white curves in
    (c) are 2D field projections of the field lines mapped on the
    slice.}
  \label{fig:recon_sample}
\end{figure*}

\begin{figure}[htbp]
  \centering
  \includegraphics[width=0.4\textwidth]{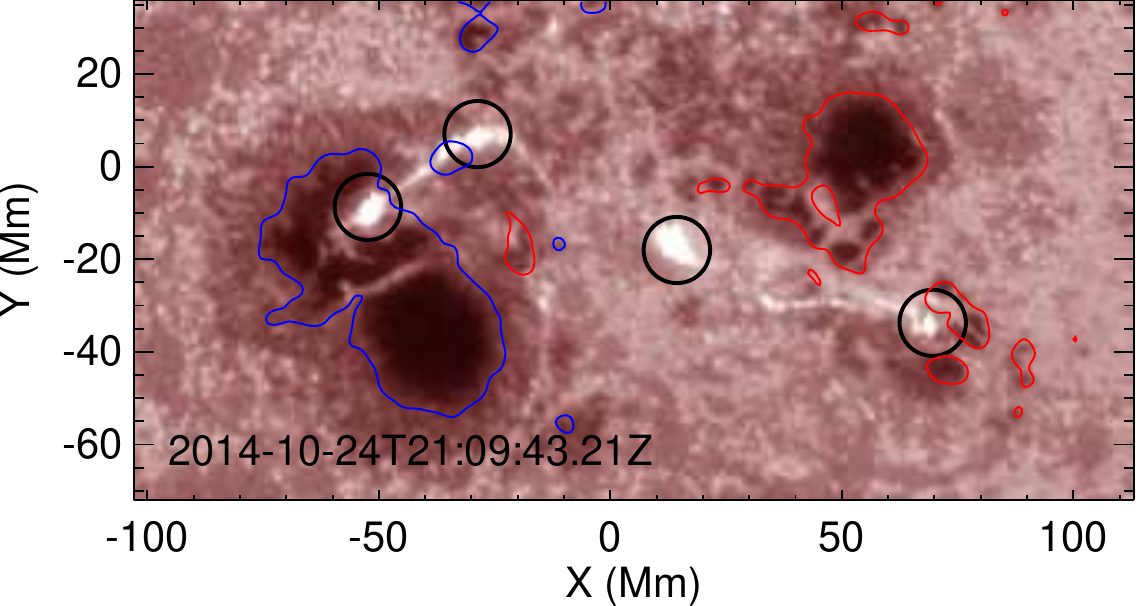}
  \caption{AIA 1700~{\AA} image of the flare ribbon at the beginning
    of the flare. Four initial brightening flare patches are marked by
    circles. The contour lines represent photospheric $B_z$ of $\pm
    1000$~G.}
  \label{fig:1700}
\end{figure}

\begin{figure}[htbp]
  \centering
  \includegraphics[width=0.4\textwidth]{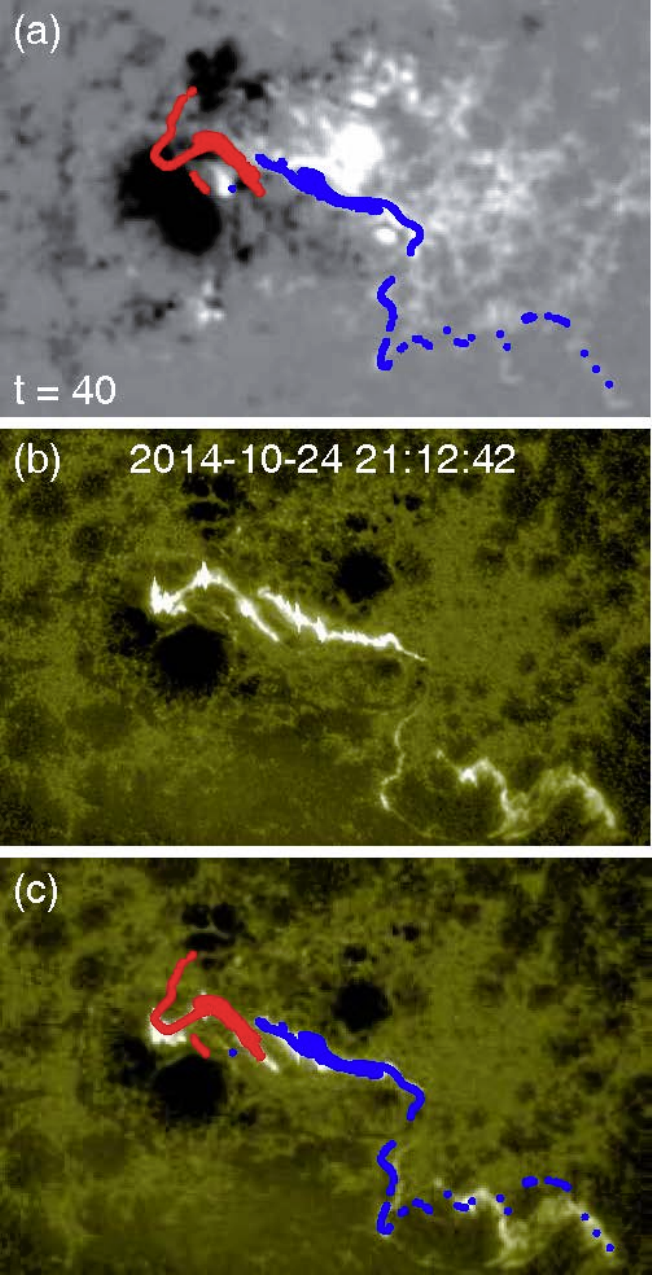}
  \caption{Comparison of the footpoints of reconnecting field lines in
    the model with the observed flare ribbons. (a) The modeled
    footpoints overlaid on photospheric magnetogram of $B_{z}$
    (saturated at $\pm 1000$~G). Results are produced at the modeling
    time $t=40$, when the CS size attained the peak value. Footpoints
    in positive (negative) magnetic flux are shown with blue (red)
    color. (b) AIA 1600~{\AA} image of the chromospheric flare
    ribbons. The image is re-mapped and co-aligned with the magnetogram
    shown in (a). (c) The modeled footpoints overplotted on the AIA
    image.}
  \label{fig:ribbon_compare}
\end{figure}

\begin{figure*}[htbp]
  \centering
  \includegraphics[width=0.8\textwidth]{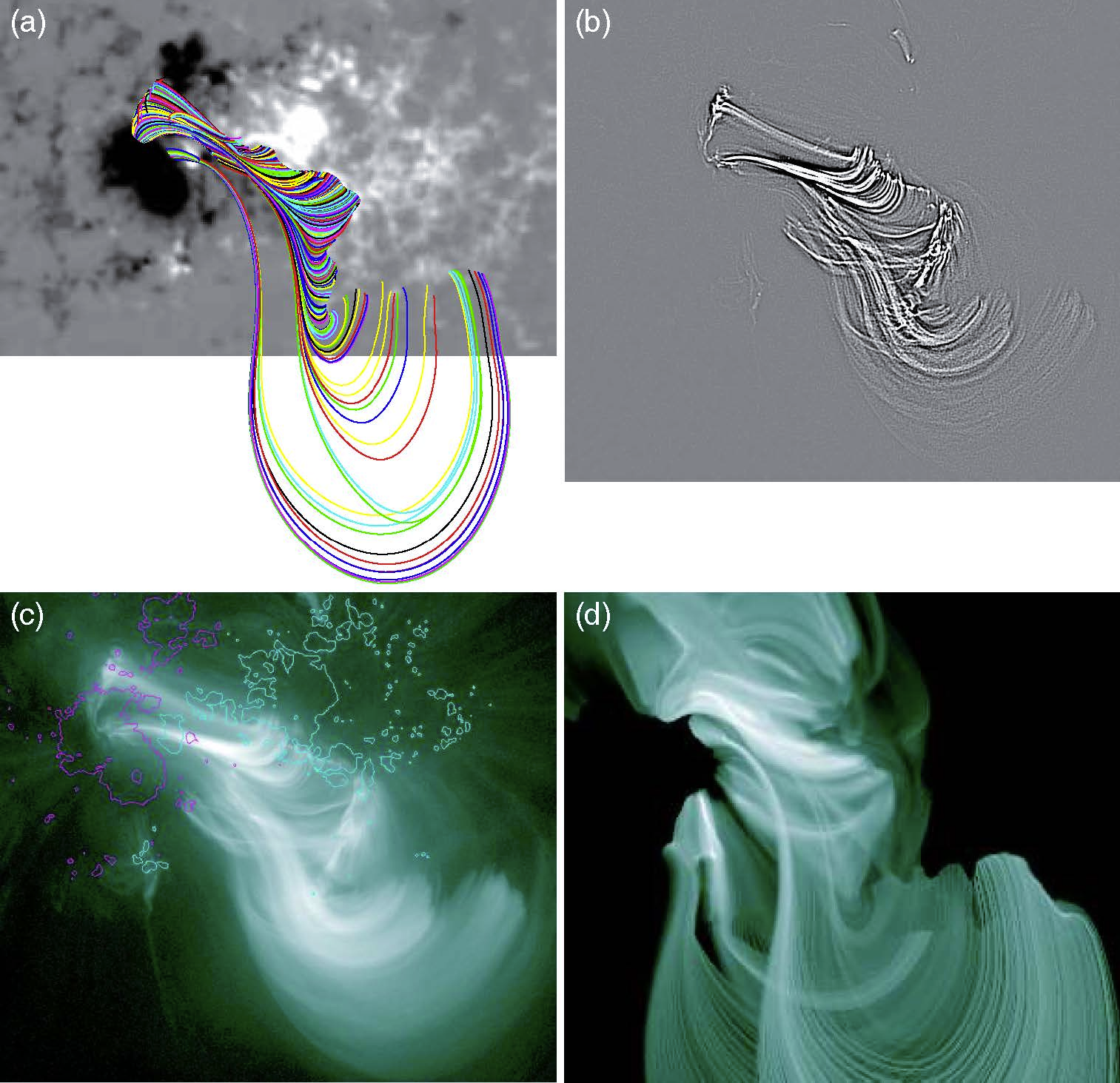}
  \caption{Comparison of the reconnecting field lines from the model
    with the observed flaring loops. (a) The reconnecting field lines
    (modeling time at $t=40$). They are shown with different colors
    for better discrimination of each field lines. (b) High-pass
    filtered AIA 94~{\AA} image highlighting the flaring hot
    loops. (c) Original AIA 94~{\AA} image. Overlaid contours
    represent photospheric $B_{z}$ of $\pm 1000$~G. (d) A Synthetic
    EUV image of the flaring loops based on the modeled magnetic field
    and currents (specified in the text). }
  \label{fig:loop_compare}
\end{figure*}

\subsection{Reconnection configuration and comparison with observations}
\label{sec:3.3}

In \Fig~\ref{fig:recon_sample}a and b we trace sampled field lines to
illustrate the reconnection configuration. These four field lines are
selected from the model at the peak time $t=40$ of reconnection. The
yellow and white lines are traced from the core site of the CS (shown
by the pink object) and they represent the pre-reconnection field
lines. These field lines (yellow and white) are sheared and pass each
other at their inner footpoints, while the CS was formed at the
interface between the crossing field lines. These field lines later
reconnected and changed their connectivity forming a longer field line
(blue) connecting the two outer-most footpoints and a shorter line
(cyan) connecting the two inner footpoints. By the magnetic tension
force, both the post-reconnection field lines relax. The longer field
line expands upward but only slightly, and the shorter line contracts
downward.  Further evidence of this type of reconnection are the
plasma inflows and outflows associated with the CS (i.e., diffusive
region), horizontal and vertical slices of which made near the
reconnection point are shown, in \Fig~\ref{fig:recon_sample}c and
d. Although the reconnection configuration are of full 3D, shown in
the vertical slice they exhibits a typical \textsf{X}-shaped
two-dimensional (2D) reconnection picture.  Thus locally such
reconnection can be considered in a 2D framework with a strong guide
field (i.e., out-of-the-plane component).

To support the association of the modeled reconnection configuration
with the real flare event, observed signatures of the flare emission
can be compared with the modeled results, in an indirect way. Among
these are the location and shape of the chromospheric flare ribbons,
which are recognized to be an indicator of the footpoint locations of
those magnetic field lines that underwent reconnection
\citep{Qiu2009}. This is mainly due to the non-thermal particles
accelerated at the reconnection site traveling down along these field
lines toward the photosphere and colliding with the dense chromosphere
causing enhanced heating~\citep{Reid2012}.

As the sampled field lines shown in \Fig~\ref{fig:recon_sample} are
traced from the core site of the CS (where $J/B$ is the strongest),
they can be regarded as the ``first'' field lines that reconnect in
the whole simulated flare process (however, it is difficult to
precisely locate the first reconnection point and field lines in the
model).  These first reconnected field lines have four footpoints at
the bottom, and each of them, from left to right as shown in
\Fig~\ref{fig:recon_sample}a, roughly corresponding to the four
brightening patches as shown in \Fig~\ref{fig:1700}, from left to
right, respectively, which are identified in AIA 1700~{\AA} channel at
the flare beginning. Both the simulation and the observation show that
the two inner flaring points are separated by a relatively large
distance of approximately $50$~Mm. The simulation also suggests that
the initial reconnection point is already rather high ($\sim 30$~Mm)
in the corona.

Following the initial reconnection, the reconnection site should
extends horizontally and forms a line of reconnection points with
similar \textsf{X}-shaped configuration at their vertical slice. Along
the central line of the CS (see the regions with $J/B > 0.3/\Delta x$
shown in \Fig~\ref{fig:recon_sample}c), we can see the horizontal
reconnection inflow vectors are almost perpendicular to the CS. This
central line is a hint of such reconnection line. When all the field
lines along the reconnection line are involved in, the two ribbons
form (see \Fig~\ref{fig:ribbon_compare}b). To simulate such ribbons,
we identify all the reconnecting field lines as those that are in
contact or pass through the CS (see \Fig~\ref{fig:loop_compare}a),
which means that these field lines were undergoing reconnection or
formed immediately after reconnection. The footpoints of these field
lines, when mapped on the photosphere, appear to form two curved
narrow areas separated by the central PIL
(\Fig~\ref{fig:ribbon_compare}a). A strikingly good match in both the
location and the shape of these simulated footpoint areas with the
observed chromospheric flare ribbons is evident
(\Fig~\ref{fig:ribbon_compare}b and c), including even the relatively
weak ribbon extending far in the southwest direction.

The flaring loop in the hot channels (e.g., AIA-94) can also be
compared with our simulated field (see \Fig~\ref{fig:loop_compare}a
and b). For a better vision comparison, we generated a synthetic EUV
image using a method similar to that used by \citet{Cheung2012}. We
first trace a large number ($3\times 10^5$) of field lines from their
footpoints uniformly distributed at the bottom. Then we assign for
each field line a value of emission assumed to be $J/B J^2$ at the
peak value point of $J/B$ along the field line. Here $J/B$ can be
regarded as the dissipation rate of current, and by selecting its
maximum, we can emphasize the emission from the reconnecting field
lines. Finally the total emission is obtained by integration through
the volume along the line-of-sight (here simply along the $z$-axis),
and forms the synthetic image. As shown in
\Fig~\ref{fig:loop_compare}c and d, a good morphological similarity is
achieved between the simulated emission and the AIA-94 images of the
hot flaring loops.

As can be seen from the comparison, the ribbons in the core region
correspond to the footpoints of the short field lines there, which
reveal themselves as the short contracting loops.  The far southwest
ribbon is due to the long field lines that connect the
negative-polarity sunspot and the far southwest plage region next to
the positive-polarity sunspot, and these long field lines correspond
to the long flaring loops.  This might explains why flares frequently
involve the brightening of these long side loops. We note that the
shape of long loops appear not to be very well reproduced. This is
possibly because they are close to the side boundaries of the
simulation volume and thus the modeling of the field there is subject
to the influence of the numerical boundary conditions, especially
during such a long-term computation. A more important factor could be
that our model is based on Cartesian geometry, thus failed to
accurately reproduce the very long loops for which the curvature of
the Sun must be considered.  Nonetheless, the mapping of the field
lines to the bottom surface is still accurate.

\section{Discussions}
\label{sec:con}

We simulated the magnetic dynamics of AR~12192 in a two-day period
using the DARE model with the SDO/HMI vector magnetograms as evolving
boundary input. Analysis of the modeled results shows that a
large-scale CS is developed in the AR core field due to stressing by
photospheric driving, and then reconnection is triggered within the
CS, resulting in an impulsive release of magnetic energy, which could
correspond to an X3.1 flare occurred near the end of the period. The
reconnection configuration exhibits signatures of the tether-cutting
reconnection~\citep{Moore2001}.  Comparison with the AIA observations
shows that the model almost reproduced exactly the location of the
chromospheric flare ribbons, and the morphology of the reconnecting
field lines and the simulated EUV image resembles well the flaring
coronal loops.  Such an agreement of simulation with observations
supports that our model correctly captured the essentials of the MHD
process of this flare.  Observations~\citep{ChenH2015} show that the
X-flares in the same AR exhibited very similar flaring structures,
indicating that these flares were homologous flares with analogous
magnetic mechanism. This might indicate multiple recurrence of the
process from slow formation to fast dissipation of the intensive
current layer between the sheared arcades, while the large-scale
configuration of the AR does not change much.

Analysis of the magnetic decay index \citep{SunX2015, Inoue2016,
  JingJ2015} seems to explain why the flare failed to erupt, as the
overlying closed magnetic field is sufficiently strong to confine the
eruption from below~\citep[see another failed eruption shown
in][]{WangH2015}.  Here the DARE simulation provides additional and
direct explanations. As can be seen from a direct look of the field
lines (see \Fig~\ref{fig:recon_sample}), the pre-flare magnetic
arcades are twisted (or stressed) by a rather weak extent. A
quantitative study of the magnetic twist has been performed by
\citet{Inoue2016} for the same flare event using a NLFFF model. They
found that the magnetic twists before the flare is mostly less than a
half turn.  Consequently, the sheared magnetic arcades do not show a
well-formed two-\textsf{J} shape like those in many sigmoid ARs with
eruptive events. \citet{Inoue2016} also showed that after the flare,
the magnetic twists are almost reserved without release.  This is
consistent with our modeled results that the post-reconnection field
lines are still sheared arcades, without forming an escaping magnetic
flux rope. In our simulation, the reconnected long field lines on top
of the CS only expands slightly without propagating further to make
the overlying arcades open. Thus, in the context of tether-cutting
scenario, only the first-stage of tether-cutting occurred, while the
second stage, formation of a flux rope and reconnection below the flux
rope, did not happen.

Another unusual fact of this flare is that the enhancements of the
horizontal field at the photosphere is very weak~\citep{SunX2015}.
This is unlike in many other large flares, as an evident enhancement
of photospheric horizontal field after flare is commonly
observed~\citep{WangH2006, WangH2010, WangS2012, Petrie2012}. In the
proposed coronal ``implosion'' mechanism~\citep{Hudson2000}, such
enhancement of photospheric field is due to the downward contraction
of the short magnetic arcades formed after the reconnection and their
push on the photosphere. As we found in the simulation here, the
reconnection site is rather high above the photosphere, which is also
suggested by \citet{Thalmann2015} from the study of the limited
variation of the flare-ribbon separations. As a result, the
reconnected short field lines below the CS expanded still rather
highly in the corona. So during its contraction, the effect of its
push on the bottom might be reduced significantly by the strong corona
field below. Thus the large altitude of the reconnection sites and the
long extension of the reconnected arcades provide a plausible
explanation for the weak ``implosion'' effect. A further quantitative
analysis of this effect will be considered in future works.

It has been found that the magnetic energies from NLFFF models usually
underestimate the related flare energies \citep[e.g.,][]{Sun2012,
  FengL2013}. For the present X3.1 flare, \citet{SunX2015} gives a
result of $0.9\times 10^{32}$~erg from a NLFFF extrapolation code. As
a reference, \citet{Thalmann2015} estimated that the non-thermal
electron energy for an earlier, confined X1 flare as $1.6\times
10^{32}$~erg, so the X3.1 flare energy is likely to be much larger
than this value. On the other hand, from our simulation we have
estimated a total released magnetic energy of $10^{33}$~erg, which is
an order higher than that derived from the NLFFF model
\citep{SunX2015} and appears to be sufficient for powering the X3.1
flare. Such an unusually large value of energy release (compared with
typical ARs) is still reasonable when considering the unusually large
size of this AR with potential energy of about $1.5\times 10^{34}$, a
order higher than typical ARs. While the NLFFF model calculates the
drop of total magnetic energy of the whole modeling volume from
pre-flare to post-flare states, we can directly calculate the magnetic
energy lost in the CS due to the ``reconnection'' within the CS, which
should be more relevant with the flare-released energy. As such, we
did not perform energy analysis for the full modeling volume.

Recently, \citet{Savcheva2015} claimed that based on MF or NLFFF
models and searching of QSLs, they were able to predict the locations
of the flare ribbon. However, we note that the QSLs are only possible
sites for reconnection, while they cannot tell where the reconnection
occurs specifically in a flare.  Moreover, by inspecting the map of
QSLs, one usually see much more complex structures than that of target
flare ribbons (see Figure~5 in \citet{Savcheva2015}). Thus without
knowing ribbon locations in advance, it is still problematic to
identify from all the QSLs the particular flare-related one. Here with
the MHD model we simulated the flare reconnection process in a
self-consistent way and directly identified the reconnecting field
lines, and thus we are able to almost reproduce precisely the flare
ribbon locations.

We find that the modeled ``flare'' began (at $t\approx 30$) well
before the photospheric magnetic field input at the bottom boundary
evolving to the time of the real flare (at $t\approx 45$).  A perfect
model of reproducing the reality should produce a flare at the exact
time when the photospheric field reaches the flare onset time. The
mismatch of our model with the reality is probably in a large part due
to the over-simplification of the magnetic reconnection process, which
might be much more complex since it is related to the microscopic
behavior of plasma. However, as in many other solar MHD codes, the
modeled reconnection here is simply resulted by numerical viscosity in
the CS region, and its behavior depends on the numerical aspects of
the model. For example, the thickness of the modeled CS and the
numerical viscosity are often sensitive to the grid resolution (as
well as the specific numerical scheme). A much thinner CS can develop
if using a smaller grid size, and the onset of reconnection in the CS
might be postponed. Namely, with a smaller grid size, a CS can sustain
even stronger current and thus even larger gradient of magnetic field,
which might needs more time to form. Further experiments using
different grid resolutions will be required to quantify this effect.

Comparative study of this flare with eruptive ones may provide insight
in the different magnetic natures of eruptive and confined flares.
Here we refer to our previous study in which we use the DARE model to
simulate an eruption event in AR~11283~\citep{Jiang2016NC}. For that
event, the eruption onset time is matched by the model much better,
with less than a time lag of $2\tau$ during a whole simulated period
of $60\tau$. Unlike this event, the simulated eruption in AR~11283 is
due to the formation of a jet-like magnetic configuration that favors
breakout-like reconnection~\citep{Antiochos1999}.  So in that event,
the critical condition causing eruption is the formation of a
reconnection-favorable magnetic topology, which can be regarded as a
macroscopic behavior. Whereas in AR~12192, cause of its flare depends
more on the triggering of the reconnection rather than the formation
of a favorable topology, since the basic topology is simply a
configuration of sheared arcades and it does not change for days. In
other words, the cause of the flare relies more on the microscopic
behavior of the plasma and thus is more subtle. The MHD model seems to
be able to characterize well the macroscopic structures and
evolutions, while it may not appropriately simulate the microscopic
aspects of the plasma, and thus, the triggering of the reconnection.
Such difference might be common and even fundamental between eruptive
flares and confined flares, in particular, those events without
noticeable changes of the coronal structures.


Finally, the reconnection-related quantities derived from the model
should be taken with cautions because of the simplification of
reconnection.  A deeper understanding of the flare dynamics requires
knowledge of the true nature of magnetic reconnection and is beyond
the scope of this paper. In addition, the quantitative results
suffered uncertainties from several aspects including HMI data and
model settings. For example, the evolution speed of the photospheric
magnetic field is increased in the model for saving computing time,
and this might affect the coronal evolution, most likely at the time
around the flare.

In summary, the present data-driven simulation study can provide
important insight in understanding why and how solar flares occur,
particularly, for those events in which the dynamic change is elusive
in observations. Further advancements, including more realistic plasma
model, reconnection realization and thermodynamics as well as
extension to spherical geometry, are necessary for even more
sophisticated modeling of real solar flares. With these improvements,
the DARE model will hopefully become a useful tool to the communities
of solar physics and space weather.

\acknowledgments

This work is supported by NSF AGS-1153323, AGS-1062050 and in addition
C.J. and X.F. are also supported by the 973 program under grant
2012CB825601, the Chinese Academy of Sciences (KZZD-EW-01-4), the
National Natural Science Foundation of China (41204126, 41231068,
41274192, 41531073, 41374176, 41574170, and 41574171), and the
Specialized Research Fund for State Key Laboratories and Youth
Innovation Promotion Association of CAS (2015122). V.Y. acknowledges
support from AFOSR FA9550-15-1-0322 and NSF AGS-1250818 grants and
Korea Astronomy and Space Science Institute. H.W. acknowledges support
from NSF AGS-1348513, AGS-1408703. Data from observations are courtesy
of NASA {SDO}/AIA and the HMI science teams. We thank the anonymous
referee for help of improving the manuscript. We thank International
Space Science Institute (ISSI) for enabling interesting discussions.


\begin{thebibliography}{88}
\expandafter\ifx\csname natexlab\endcsname\relax\def\natexlab#1{#1}\fi

\bibitem[{{Amari} {et~al.}(1997){Amari}, {Aly}, {Luciani}, {Boulmezaoud}, \&
  {Mikic}}]{Amari1997}
{Amari}, T., {Aly}, J.~J., {Luciani}, J.~F., {Boulmezaoud}, T.~Z., \& {Mikic},
  Z. 1997, \solphys, 174, 129

\bibitem[{Amari {et~al.}(2014)Amari, Canou, \& Aly}]{Amari2014nat}
Amari, T., Canou, A., \& Aly, J.~J. 2014, Nature, 514, 465

\bibitem[{{Antiochos} {et~al.}(1999){Antiochos}, {DeVore}, \&
  {Klimchuk}}]{Antiochos1999}
{Antiochos}, S.~K., {DeVore}, C.~R., \& {Klimchuk}, J.~A. 1999, \apj, 510, 485

\bibitem[{{Aulanier} {et~al.}(2007){Aulanier}, {Golub}, {DeLuca}, {Cirtain},
  {Kano}, {Lundquist}, {Narukage}, {Sakao}, \& {Weber}}]{Aulanier2007}
{Aulanier}, G., {Golub}, L., {DeLuca}, E.~E., {Cirtain}, J.~W., {Kano}, R.,
  {Lundquist}, L.~L., {Narukage}, N., {Sakao}, T., \& {Weber}, M.~A. 2007,
  Science, 318, 1588

\bibitem[{{Bobra} {et~al.}(2014){Bobra}, {Sun}, {Hoeksema}, {Turmon}, {Liu},
  {Hayashi}, {Barnes}, \& {Leka}}]{Bobra2014}
{Bobra}, M.~G., {Sun}, X., {Hoeksema}, J.~T., {Turmon}, M., {Liu}, Y.,
  {Hayashi}, K., {Barnes}, G., \& {Leka}, K.~D. 2014, \solphys, 289, 3549

\bibitem[{{Carmichael}(1964)}]{Carmichael1964}
{Carmichael}, H. 1964, NASA Special Publication, 50, 451

\bibitem[{Chen {et~al.}(2015)Chen, Zhang, Ma, Yang, Li, Huang, \&
  Xiao}]{ChenH2015}
Chen, H., Zhang, J., Ma, S., Yang, S., Li, L., Huang, X., \& Xiao, J. 2015,
  \apjl, 808, L24

\bibitem[{{Cheung} \& {DeRosa}(2012)}]{Cheung2012}
{Cheung}, M.~C.~M. \& {DeRosa}, M.~L. 2012, \apj, 757, 147

\bibitem[{Courant {et~al.}(1967)Courant, Friedrichs, \&
  Lewy}]{courant1967partial}
Courant, R., Friedrichs, K., \& Lewy, H. 1967, IBM journal of Research and
  Development, 11, 215

\bibitem[{{Dai} {et~al.}(2013){Dai}, {Ding}, \& {Guo}}]{Dai2013}
{Dai}, Y., {Ding}, M.~D., \& {Guo}, Y. 2013, \apjl, 773, L21

\bibitem[{{Dalmasse} {et~al.}(2015){Dalmasse}, {Chandra}, {Schmieder}, \&
  {Aulanier}}]{Dalmasse2015}
{Dalmasse}, K., {Chandra}, R., {Schmieder}, B., \& {Aulanier}, G. 2015, \aap,
  574, A37

\bibitem[{{Demoulin} {et~al.}(1996){Demoulin}, {Henoux}, {Priest}, \&
  {Mandrini}}]{Demoulin1996}
{Demoulin}, P., {Henoux}, J.~C., {Priest}, E.~R., \& {Mandrini}, C.~H. 1996,
  \aap, 308, 643

\bibitem[{Dud{\'\i}k {et~al.}(2016)Dud{\'\i}k, Polito, Janvier, Mulay,
  Karlick{\`y}, Aulanier, Del~Zanna, Dzif{\v{c}}{\'a}kov{\'a}, Mason, \&
  Schmieder}]{Dudik2016}
Dud{\'\i}k, J., Polito, V., Janvier, M., Mulay, S.~M., Karlick{\`y}, M.,
  Aulanier, G., Del~Zanna, G., Dzif{\v{c}}{\'a}kov{\'a}, E., Mason, H.~E., \&
  Schmieder, B. 2016, The Astrophysical Journal, 823, 41

\bibitem[{{Feng} {et~al.}(2013){Feng}, {Wiegelmann}, {Su}, {Inhester}, {Li},
  {Sun}, \& {Gan}}]{FengL2013}
{Feng}, L., {Wiegelmann}, T., {Su}, Y., {Inhester}, B., {Li}, Y.~P., {Sun},
  X.~D., \& {Gan}, W.~Q. 2013, \apj, 765, 37

\bibitem[{Fisher {et~al.}(2015)Fisher, Abbett, Bercik, Kazachenko, Lynch,
  Welsch, Hoeksema, Hayashi, Liu, Norton, Dalda, Sun, DeRosa, \&
  Cheung}]{Fisher2015}
Fisher, G.~H., Abbett, W.~P., Bercik, D.~J., Kazachenko, M.~D., Lynch, B.~J.,
  Welsch, B.~T., Hoeksema, J.~T., Hayashi, K., Liu, Y., Norton, A.~A., Dalda,
  A.~S., Sun, X., DeRosa, M.~L., \& Cheung, M. C.~M. 2015, Space Weather, 13,
  369, 2015SW001191

\bibitem[{{Forbes} {et~al.}(2006){Forbes}, {Linker}, {Chen}, {Cid}, {K{\'o}ta},
  {Lee}, {Mann}, {Miki{\'c}}, {Potgieter}, {Schmidt}, {Siscoe}, {Vainio},
  {Antiochos}, \& {Riley}}]{Forbes2006}
{Forbes}, T.~G., {Linker}, J.~A., {Chen}, J., {Cid}, C., {K{\'o}ta}, J., {Lee},
  M.~A., {Mann}, G., {Miki{\'c}}, Z., {Potgieter}, M.~S., {Schmidt}, J.~M.,
  {Siscoe}, G.~L., {Vainio}, R., {Antiochos}, S.~K., \& {Riley}, P. 2006, \ssr,
  123, 251

\bibitem[{Gibson \& Fan(2006)}]{GibsonFan2006}
Gibson, S.~E. \& Fan, Y. 2006, J. Geophys. Res., 111, A12103

\bibitem[{Gou {et~al.}(2016)Gou, Liu, Wang, Liu, Zhuang, Chen, Zhang, \&
  Liu}]{GouT2016}
Gou, T., Liu, R., Wang, Y., Liu, K., Zhuang, B., Chen, J., Zhang, Q., \& Liu,
  J. 2016, The Astrophysical Journal Letters, 821, L28

\bibitem[{Grad \& Rubin(1958)}]{Grad1958}
Grad, H. \& Rubin, H. 1958, in 2nd Int. Conf. Peac. Uses of Atom. Energy,
  Vol.~31, 386

\bibitem[{{Guo} {et~al.}(2010){Guo}, {Ding}, {Schmieder}, {Li},
  {T{\"o}r{\"o}k}, \& {Wiegelmann}}]{GuoY2010}
{Guo}, Y., {Ding}, M.~D., {Schmieder}, B., {Li}, H., {T{\"o}r{\"o}k}, T., \&
  {Wiegelmann}, T. 2010, \apjl, 725, L38

\bibitem[{{He} \& {Wang}(2006)}]{He2006}
{He}, H. \& {Wang}, H. 2006, \mnras, 369, 207

\bibitem[{{Hirayama}(1974)}]{Hirayama1974}
{Hirayama}, T. 1974, \solphys, 34, 323

\bibitem[{{Hoeksema} {et~al.}(2014){Hoeksema}, {Liu}, {Hayashi}, {Sun},
  {Schou}, {Couvidat}, {Norton}, {Bobra}, {Centeno}, {Leka}, {Barnes}, \&
  {Turmon}}]{Hoeksema2014}
{Hoeksema}, J.~T., {Liu}, Y., {Hayashi}, K., {Sun}, X., {Schou}, J.,
  {Couvidat}, S., {Norton}, A., {Bobra}, M., {Centeno}, R., {Leka}, K.~D.,
  {Barnes}, G., \& {Turmon}, M. 2014, \solphys, 289, 3483

\bibitem[{Hudson(2000)}]{Hudson2000}
Hudson, H.~S. 2000, The Astrophysical Journal Letters, 531, L75

\bibitem[{{Inoue} {et~al.}(2016){Inoue}, {Hayashi}, \& {Kusano}}]{Inoue2016}
{Inoue}, S., {Hayashi}, K., \& {Kusano}, K. 2016, \apj, 818, 168

\bibitem[{Inoue {et~al.}(2014)Inoue, Hayashi, Magara, Choe, \&
  Park}]{Inoue2014}
Inoue, S., Hayashi, K., Magara, T., Choe, G.~S., \& Park, Y.~D. 2014, The
  Astrophysical Journal, 788, 182

\bibitem[{Inoue {et~al.}(2015)Inoue, Hayashi, Magara, Choe, \&
  Park}]{Inoue2015}
---. 2015, The Astrophysical Journal, 803, 73

\bibitem[{Janvier {et~al.}(2016)Janvier, Savcheva, Pariat, Tassev, Millholland,
  Bommier, McCauley, McKillop, \& Dougan}]{Janvier2016}
Janvier, M., Savcheva, A., Pariat, E., Tassev, S., Millholland, S., Bommier,
  V., McCauley, P., McKillop, S., \& Dougan, F. 2016, arXiv preprint
  arXiv:1604.07241

\bibitem[{{Ji} {et~al.}(2003){Ji}, {Wang}, {Schmahl}, {Moon}, \&
  {Jiang}}]{Ji2003}
{Ji}, H., {Wang}, H., {Schmahl}, E.~J., {Moon}, Y.-J., \& {Jiang}, Y. 2003,
  \apjl, 595, L135

\bibitem[{{Jiang} \& {Feng}(2013)}]{Jiang2013NLFFF}
{Jiang}, C. \& {Feng}, X. 2013, \apj, 769, 144

\bibitem[{{Jiang} {et~al.}(2012){Jiang}, {Feng}, {Wu}, \& {Hu}}]{Jiang2012c}
{Jiang}, C., {Feng}, X., {Wu}, S.~T., \& {Hu}, Q. 2012, \apj, 759, 85

\bibitem[{{Jiang} \& {Feng}(2012)}]{Jiang2012apj}
{Jiang}, C.~W. \& {Feng}, X.~S. 2012, \apj, 749, 135

\bibitem[{Jiang {et~al.}(2013)Jiang, Feng, Wu, \& Hu}]{Jiang2013MHD}
Jiang, C.~W., Feng, X.~S., Wu, S.~T., \& Hu, Q. 2013, \apjl, 771, L30

\bibitem[{Jiang {et~al.}(2010)Jiang, Feng, Zhang, \& Zhong}]{Jiang2010}
Jiang, C.~W., Feng, X.~S., Zhang, J., \& Zhong, D.~K. 2010, \solphys, 267, 463

\bibitem[{{Jiang} {et~al.}(2016){Jiang}, {Wu}, {Feng}, \& {Hu}}]{Jiang2016NC}
{Jiang}, C.~W., {Wu}, S.~T., {Feng}, X.~S., \& {Hu}, Q. 2016, Nature Comm., 7,
  11522

\bibitem[{Jing {et~al.}(2015)Jing, Xu, Lee, Nitta, Liu, Park, Wiegelmann, \&
  Wang}]{JingJ2015}
Jing, J., Xu, Y., Lee, J., Nitta, N.~V., Liu, C., Park, S.~H., Wiegelmann, T.,
  \& Wang, H.~M. 2015, Research in Astronomy and Astrophysics, 15, 1537

\bibitem[{{Kliem} {et~al.}(2013){Kliem}, {Su}, {van Ballegooijen}, \&
  {DeLuca}}]{Kliem2013}
{Kliem}, B., {Su}, Y.~N., {van Ballegooijen}, A.~A., \& {DeLuca}, E.~E. 2013,
  \apj, 779, 129

\bibitem[{{Kopp} \& {Pneuman}(1976)}]{Kopp1976}
{Kopp}, R.~A. \& {Pneuman}, G.~W. 1976, \solphys, 50, 85

\bibitem[{{Li} \& {Zhang}(2013)}]{LiT2013a}
{Li}, T. \& {Zhang}, J. 2013, \apjl, 778, L29

\bibitem[{{Li} \& {Zhang}(2015)}]{LiT2015}
---. 2015, \apjl, 804, L8

\bibitem[{Liu {et~al.}(2009)Liu, Lee, Karlicky, Choudhary, Deng, \&
  Wang}]{LiuC2009}
Liu, C., Lee, J., Karlicky, M., Choudhary, D.~P., Deng, N., \& Wang, H. 2009,
  The Astrophysical Journal, 703, 757

\bibitem[{Liu {et~al.}(2013)Liu, Zhang, Wang, \& Cheng}]{LiuK2013}
Liu, K., Zhang, J., Wang, Y., \& Cheng, X. 2013, The Astrophysical Journal,
  768, 150

\bibitem[{{Liu} {et~al.}(2012){Liu}, {Zhao}, \& {Schuck}}]{Liu2012}
{Liu}, Y., {Zhao}, J., \& {Schuck}, P.~W. 2012, \solphys, 195

\bibitem[{{Masson} {et~al.}(2009){Masson}, {Pariat}, {Aulanier}, \&
  {Schrijver}}]{Masson2009}
{Masson}, S., {Pariat}, E., {Aulanier}, G., \& {Schrijver}, C.~J. 2009, \apj,
  700, 559

\bibitem[{{Moore} {et~al.}(2001){Moore}, {Sterling}, {Hudson}, \&
  {Lemen}}]{Moore2001}
{Moore}, R.~L., {Sterling}, A.~C., {Hudson}, H.~S., \& {Lemen}, J.~R. 2001,
  \apj, 552, 833

\bibitem[{{Nakagawa} {et~al.}(1987){Nakagawa}, {Hu}, \& {Wu}}]{Nakagawa1987}
{Nakagawa}, Y., {Hu}, Y.~Q., \& {Wu}, S.~T. 1987, \aap, 179, 354

\bibitem[{{Petrie}(2012)}]{Petrie2012}
{Petrie}, G.~J.~D. 2012, \apj, 759, 50

\bibitem[{{Priest} \& {Forbes}(2002)}]{Priest2002}
{Priest}, E.~R. \& {Forbes}, T.~G. 2002, \aapr, 10, 313

\bibitem[{{Qiu}(2009)}]{Qiu2009}
{Qiu}, J. 2009, \apj, 692, 1110

\bibitem[{R{\'e}gnier(2013)}]{Regnier2013}
R{\'e}gnier, S. 2013, Solar Physics, 288, 481

\bibitem[{{Reid} {et~al.}(2012){Reid}, {Vilmer}, {Aulanier}, \&
  {Pariat}}]{Reid2012}
{Reid}, H.~A.~S., {Vilmer}, N., {Aulanier}, G., \& {Pariat}, E. 2012, \aap,
  547, A52

\bibitem[{{Romano} {et~al.}(2015){Romano}, {Zuccarello}, {Guglielmino},
  {Berrilli}, {Bruno}, {Carbone}, {Consolini}, {de Lauretis}, {Del Moro},
  {Elmhamdi}, {Ermolli}, {Fineschi}, {Francia}, {Kordi}, {Landi
  Degl'Innocenti}, {Laurenza}, {Lepreti}, {Marcucci}, {Pallocchia},
  {Pietropaolo}, {Romoli}, {Vecchio}, {Vellante}, \& {Villante}}]{Romano2015}
{Romano}, P., {Zuccarello}, F., {Guglielmino}, S.~L., {Berrilli}, F., {Bruno},
  R., {Carbone}, V., {Consolini}, G., {de Lauretis}, M., {Del Moro}, D.,
  {Elmhamdi}, A., {Ermolli}, I., {Fineschi}, S., {Francia}, P., {Kordi}, A.~S.,
  {Landi Degl'Innocenti}, E., {Laurenza}, M., {Lepreti}, F., {Marcucci}, M.~F.,
  {Pallocchia}, G., {Pietropaolo}, E., {Romoli}, M., {Vecchio}, A., {Vellante},
  M., \& {Villante}, U. 2015, \aap, 582, A55

\bibitem[{Sakurai(1981)}]{Sakurai1981}
Sakurai, T. 1981, Solar physics, 69, 343

\bibitem[{Savcheva {et~al.}(2016)Savcheva, Pariat, McKillop, McCauley, Hanson,
  Su, \& DeLuca}]{Savcheva2016}
Savcheva, A., Pariat, E., McKillop, S., McCauley, P., Hanson, E., Su, Y., \&
  DeLuca, E.~E. 2016, The Astrophysical Journal, 817, 43

\bibitem[{{Savcheva} {et~al.}(2015){Savcheva}, {McKillop}, {McCauley},
  {Hanson}, {Werner}, \& {DeLuca}}]{Savcheva2015}
{Savcheva}, A.~S., {McKillop}, S.~C., {McCauley}, P.~I., {Hanson}, E.~M.,
  {Werner}, E., \& {DeLuca}, E.~E. 2015, \apj, 810, 96

\bibitem[{{Schou} {et~al.}(2012){Schou}, {Scherrer}, {Bush}, {Wachter},
  {Couvidat}, {Rabello-Soares}, {Bogart}, {Hoeksema}, {Liu}, {Duvall}, {Akin},
  {Allard}, {Miles}, {Rairden}, {Shine}, {Tarbell}, {Title}, {Wolfson},
  {Elmore}, {Norton}, \& {Tomczyk}}]{Schou2012HMI}
{Schou}, J., {Scherrer}, P.~H., {Bush}, R.~I., {Wachter}, R., {Couvidat}, S.,
  {Rabello-Soares}, M.~C., {Bogart}, R.~S., {Hoeksema}, J.~T., {Liu}, Y.,
  {Duvall}, T.~L., {Akin}, D.~J., {Allard}, B.~A., {Miles}, J.~W., {Rairden},
  R., {Shine}, R.~A., {Tarbell}, T.~D., {Title}, A.~M., {Wolfson}, C.~J.,
  {Elmore}, D.~F., {Norton}, A.~A., \& {Tomczyk}, S. 2012, \solphys, 275, 229

\bibitem[{Schrijver \& Title(2011)}]{Schrijver2011a}
Schrijver, C.~J. \& Title, A.~M. 2011, Journal of Geophysical Research: Space
  Physics, 116, 2156

\bibitem[{Schuck(2008)}]{Schuck2008}
Schuck, P.~W. 2008, The Astrophysical Journal, 683, 1134

\bibitem[{Shen {et~al.}(2012)Shen, Liu, \& Su}]{ShenY2012}
Shen, Y., Liu, Y., \& Su, J. 2012, The Astrophysical Journal, 750, 12

\bibitem[{{Shibata} \& {Magara}(2011)}]{Shibata2011}
{Shibata}, K. \& {Magara}, T. 2011, Living Reviews in Solar Physics, 8, 6

\bibitem[{{Sturrock}(1966)}]{Sturrock1966}
{Sturrock}, P.~A. 1966, \nat, 211, 695

\bibitem[{Sun {et~al.}(2015)Sun, Bobra, Hoeksema, Liu, Li, Shen, Couvidat,
  Norton, \& Fisher}]{SunX2015}
Sun, X., Bobra, M.~G., Hoeksema, J.~T., Liu, Y., Li, Y., Shen, C., Couvidat,
  S., Norton, A.~A., \& Fisher, G.~H. 2015, \apjl, 804, L28

\bibitem[{{Sun} {et~al.}(2012){Sun}, {Hoeksema}, {Liu}, {Wiegelmann},
  {Hayashi}, {Chen}, \& {Thalmann}}]{Sun2012}
{Sun}, X., {Hoeksema}, J.~T., {Liu}, Y., {Wiegelmann}, T., {Hayashi}, K.,
  {Chen}, Q., \& {Thalmann}, J. 2012, \apj, 748, 77

\bibitem[{Thalmann {et~al.}(2015)Thalmann, Su, Temmer, \&
  Veronig}]{Thalmann2015}
Thalmann, J.~K., Su, Y., Temmer, M., \& Veronig, A.~M. 2015, \apjl, 801, L23

\bibitem[{{Titov} {et~al.}(2002){Titov}, {Hornig}, \&
  {D{\'e}moulin}}]{Titov2002}
{Titov}, V.~S., {Hornig}, G., \& {D{\'e}moulin}, P. 2002, \jgr, 107, 1164

\bibitem[{{Titov} {et~al.}(1993){Titov}, {Priest}, \& {Demoulin}}]{Titov1993}
{Titov}, V.~S., {Priest}, E.~R., \& {Demoulin}, P. 1993, \aap, 276, 564

\bibitem[{{T{\"o}r{\"o}k} \& {Kliem}(2005)}]{Torok2005}
{T{\"o}r{\"o}k}, T. \& {Kliem}, B. 2005, \apjl, 630, L97

\bibitem[{{Valori} {et~al.}(2010){Valori}, {Kliem}, {T{\"o}r{\"o}k}, \&
  {Titov}}]{Valori2010}
{Valori}, G., {Kliem}, B., {T{\"o}r{\"o}k}, T., \& {Titov}, V.~S. 2010, \aap,
  519, A44+

\bibitem[{{Wang} {et~al.}(2008){Wang}, {Wu}, {Liu}, \&
  {Hathaway}}]{Wang2008ApJ}
{Wang}, A.~H., {Wu}, S.~T., {Liu}, Y., \& {Hathaway}, D. 2008, \apjl, 674, L57

\bibitem[{{Wang}(2006)}]{WangH2006}
{Wang}, H. 2006, \apj, 649, 490

\bibitem[{{Wang} {et~al.}(2015){Wang}, {Cao}, {Liu}, {Xu}, {Liu}, {Zeng},
  {Chae}, \& {Ji}}]{WangH2015}
{Wang}, H., {Cao}, W., {Liu}, C., {Xu}, Y., {Liu}, R., {Zeng}, Z., {Chae}, J.,
  \& {Ji}, H. 2015, Nature Communications, 6, 7008

\bibitem[{{Wang} \& {Liu}(2010)}]{WangH2010}
{Wang}, H. \& {Liu}, C. 2010, \apjl, 716, L195

\bibitem[{{Wang} \& {Liu}(2012)}]{WangH2012}
---. 2012, \apj, 760, 101

\bibitem[{{Wang} {et~al.}(2014){Wang}, {Liu}, {Deng}, {Zeng}, {Xu}, {Jing}, \&
  {Cao}}]{WangH2014}
{Wang}, H., {Liu}, C., {Deng}, N., {Zeng}, Z., {Xu}, Y., {Jing}, J., \& {Cao},
  W. 2014, \apjl, 781, L23

\bibitem[{{Wang} {et~al.}(2012){Wang}, {Liu}, {Liu}, {Deng}, {Liu}, \&
  {Wang}}]{WangS2012}
{Wang}, S., {Liu}, C., {Liu}, R., {Deng}, N., {Liu}, Y., \& {Wang}, H. 2012,
  \apjl, 745, L17

\bibitem[{{Welsch} {et~al.}(2004){Welsch}, {Fisher}, {Abbett}, \&
  {Regnier}}]{Welsch2004}
{Welsch}, B.~T., {Fisher}, G.~H., {Abbett}, W.~P., \& {Regnier}, S. 2004, \apj,
  610, 1148

\bibitem[{{Wheatland}(2006)}]{Wheatland2006}
{Wheatland}, M.~S. 2006, \solphys, 238, 29

\bibitem[{{Wiegelmann} \& {Neukirch}(2006)}]{Wiegelmann2006}
{Wiegelmann}, T. \& {Neukirch}, T. 2006, \aap, 457, 1053

\bibitem[{{Wiegelmann} \& {Sakurai}(2012)}]{Wiegelmann2012solar}
{Wiegelmann}, T. \& {Sakurai}, T. 2012, Living Reviews in Solar Physics, 9, 5

\bibitem[{{Woods} {et~al.}(2011){Woods}, {Hock}, {Eparvier}, {Jones},
  {Chamberlin}, {Klimchuk}, {Didkovsky}, {Judge}, {Mariska}, {Warren},
  {Schrijver}, {Webb}, {Bailey}, \& {Tobiska}}]{Woods2011}
{Woods}, T.~N., {Hock}, R., {Eparvier}, F., {Jones}, A.~R., {Chamberlin},
  P.~C., {Klimchuk}, J.~A., {Didkovsky}, L., {Judge}, D., {Mariska}, J.,
  {Warren}, H., {Schrijver}, C.~J., {Webb}, D.~F., {Bailey}, S., \& {Tobiska},
  W.~K. 2011, \apj, 739, 59

\bibitem[{{Wu} {et~al.}(1990){Wu}, {Sun}, {Chang}, {Hagyard}, \&
  {Gary}}]{Wu1990}
{Wu}, S.~T., {Sun}, M.~T., {Chang}, H.~M., {Hagyard}, M.~J., \& {Gary}, G.~A.
  1990, \apj, 362, 698

\bibitem[{{Wu} {et~al.}(2006){Wu}, {Wang}, {Liu}, \& {Hoeksema}}]{Wu2006}
{Wu}, S.~T., {Wang}, A.~H., {Liu}, Y., \& {Hoeksema}, J.~T. 2006, \apj, 652,
  800

\bibitem[{{Yamamoto} \& {Kusano}(2012)}]{Yamamoto2012}
{Yamamoto}, T.~T. \& {Kusano}, K. 2012, \apj, 752, 126

\bibitem[{{Yan} \& {Sakurai}(2000)}]{Yan2000}
{Yan}, Y. \& {Sakurai}, T. 2000, \solphys, 195, 89

\bibitem[{{Yang} {et~al.}(1986){Yang}, {Sturrock}, \& {Antiochos}}]{Yang1986}
{Yang}, W.~H., {Sturrock}, P.~A., \& {Antiochos}, S.~K. 1986, \apj, 309, 383

\bibitem[{{Yashiro} {et~al.}(2006){Yashiro}, {Akiyama}, {Gopalswamy}, \&
  {Howard}}]{Yashiro2006}
{Yashiro}, S., {Akiyama}, S., {Gopalswamy}, N., \& {Howard}, R.~A. 2006, \apjl,
  650, L143

\bibitem[{Yeates(2014)}]{Yeates2014}
Yeates, A.~R. 2014, Solar Physics, 289, 631

\bibitem[{Zhang {et~al.}(2014)Zhang, Li, \& Yang}]{ZhangJ2014}
Zhang, J., Li, T., \& Yang, S. 2014, The Astrophysical Journal Letters, 782,
  L27

\end{thebibliography}

\end{CJK*}
\end{document}